\documentclass[twocolumn,showpacs,preprintnumbers,amsmath,amssymb]{revtex4}
\usepackage{graphicx}
\usepackage{dcolumn}
\usepackage{bm}

\newcommand{\aap}{Astronom. and Astrophys.}
\newcommand{\pasa}{Publications of the Astronomical Society of Australia}

\newcommand{\aaps}{Astronom. and Astrophys. Suppl. Ser.}

\newcommand{\aj}{Astronom. J.}

\newcommand{\apss}{Astrophys. and Space Sci.}

\newcommand{\mnras}{Monthly Notices Roy. Astronom. Soc.}

\newcommand{\pasp}{Publ. Astronom. Soc. Pacific}

\newcommand{\rmxaa}{Revista Mexicana de Astronom. y Astrof.}

\def\arcsec{\hbox{$^{\prime\prime}$}}

\def\fm{\hbox{$.\!\!^m$}}
\def\fs{\hbox{$.\!\!^s$}}
\def\degr{\hbox{$^\circ$}}

\begin{document}

\title{The visually close binary system HD375; Is it a sub-giant binary?}
\author{\firstname{M.\,A.}~\surname{Al-Wardat}}
\email{mwardat@ahu.edu.jo}
\affiliation{Department of Physics, Al-Hussein Bin Talal University, P.O.Box 20, 71111, Ma'an, Jordan.}
\author{\firstname{Yu.\, Yu.}~\surname{ Balega}}
\email{balega@sao.ru}
\affiliation{Special Astrophysical Observatory of the Russian AS, Nizhnij Arkhyz, 369167 Russia}
\author{\firstname{V.\, V.}~\surname{ Leushion}}
\affiliation{Special Astrophysical Observatory of the Russian AS, Nizhnij Arkhyz, 369167 Russia}
\author{Ali Taani}
\affiliation{Applied Science Department, Aqaba University College , Al-Balqa' Applied University,  P.O. Box 1199, 77110 Aqaba, Jordan}
\author{\firstname{N.\,  A.}~\surname{ Yusuf}}
\affiliation{Physics Department, Yarmouk University, P.O.B. 566 Irbid, 21163 Jordan}
\author{\firstname{K.\, S.} ~\surname{Al-Waqfi}}
\affiliation{Physics Department, Yarmouk University, P.O.B. 566 Irbid, 21163 Jordan}
\author{\firstname{S.}~\surname{Masda}}
\affiliation{Physics Department, Yarmouk University, P.O.B. 566 Irbid, 21163 Jordan}

\received{ , 2013}%
\revised{}%

\date{\today}

\begin{abstract}
Atmospheric modeling is used to build  synthetic spectral energy distributions (SEDs) for the individual components of the speckle interferometric binary system HD375. These synthetic SEDs  are combined together for the entire system and compared with its observational SED in an iterated procedure to achieve the best fit. Kurucz blanketed models with the measurements of magnitude differences were used to build these SED's.  The input physical elements for building these best fitted synthetic SEDs represent adequately  enough the  elements of the system.
 These  elements are: $T_{\rm eff}^{a}
=6100\pm50$\,K, $T_{\rm eff}^{b} =5940\pm50$\,K, log $g_{a}=4.01\pm0.10$,
 log $g_{b}=3.98\pm0.10$, $R_a=1.93\pm0.20 R_\odot$,  $R_b=1.83\pm0.20 R_\odot$ $M_{v}^{\rm a}=3\fm26\pm0.40$, $M_{v}^{\rm b}=3\fm51\pm0.50$, $L_a= 4.63\pm0.80 L_\odot$ and $ L_b= 3.74\pm0.70 L_\odot$ depending on new estimated parallax $\pi=12.02 \pm 0.60$ mas.  A modified orbit of the system is built and compared with earlier orbits and the masses of the two components are calculated as $M_a =1.35M_{\odot}$ and $M_b=1.25M_{\odot}$.   Depending on the estimated physical and geometrical elements of the system, which are assured by synthetic photometry, we suggest that the two components are evolved subgiant (F8.5 IV \& G0 IV) stars with age of 3.5 Gy formed by fragmentation.

\end{abstract}


\maketitle
\section{Introduction}

Hipparcos mission revealed that many previously known  single stars are actually binary or multiple systems \citep{1997yCat.1239....0E}. Most  of these resolved systems are nearby stars that appear as a single star even with the largest ground-based telescopes except when  observed using high resolution techniques like speckle interferometry (SI) \citep{2002aaa...385...87B, 2010AJ....139..743T} and adaptive optics (AO) \citep{2005AJ....130.2262R, 2011MNRAS.413.1200R}. That is why these binaries took their names (Speckle Interferometric Binaries SIBs), and are also known as visually close binary systems (VCBSs).

In general, the study of binary stars is the most powerful direct method to correlate  stellar theoretical models with the actual observational elements,  which is more complicated in the case of VCBSs. It connects mass determinations with other important elements such as radius, luminosity, and effective temperature and gives a basic check of stellar structure and evolution theory \citep{2008AJ....136..312H}. It also gives a unique way for a thorough investigation of the spectral types and luminosity classes \citep{1955PASP...67..315E}.
Hundreds of such systems with periods in the order of 10 years or less, are routinely observed and analyzed by the aforementioned  high resolution techniques. But, in spite of that, there is still a paucity  in  the individual physical elements of the systems' components.
The only way to estimate these elements is by indirect analysis of the binaries. A method that makes use of Kurucz blanketed models \citep{1994KurCD..19.....K} to build a synthetic spectral energy distribution (SED) for each component separately, and hence for the entire system. Then, by comparing this entire synthetic SED  with the  observational one in an iterated repetition to achieve the best fit between them, one may be able to determine the physical and geometrical elements of the individual components.


 The method at first used earlier versions of line-blanketed plane-parallel theoretical model atmospheres for F, G, and K-type stars \citep{1988IAUS..132..531B}, where it counted  only for the hydrogen lines opacities in building the SEDs \citep{2003PhDT.......174G}.  After that, it employed ATLAS9  with its new  opacity distribution functions (ODFs)   \citep{2004astro.ph..5087C} to build the individual synthetic SEDs, and it was successfully applied to some binary systems like  Cou1289, Cou1291, Hip11352, Hip11253, Hip70973 and Hip72479  \citep{2007AN....328...63A, 2009AN....330..385A, 2009AstBu..64..365A, 2012PASA...29..523A}.

 The VCBS   HD375 was firstly analyzed using the earlier version of this method by \cite{2003PhDT.......174G}.  The modified physical and geometrical elements for the system using the modified version of the aforementioned method, and the modified orbit of the system depending on latest SI observations are presented.   These information will  enhance our knowledge about stellar parameters in general and consequently will help  in understanding the formation
and evolution mechanisms of stellar binary  systems.

\section{Atmospheric modeling}

Table~\ref{table1} contains basic data of the system from SIMBAD, NASA/IPAC and Table~\ref{table2}
contains data from Hipparcos and Tycho Catalogues \citep{1997yCat.1239....0E}.

\begin{table}
\begin{center}
\caption{Data from SIMBAD and NASA/IPAC.} \label{table1}
\begin{tabular}{l|c|c}\hline
  & Hip689  & ref.
   \\
   & HD375  &
   \\
   & HDS17& \\
 \hline
$\alpha_{2000}$ & $00^h 08^m 28\fs446$ & 1
\\
$\delta_{2000}$&$+34\degr56' 04.''37$ & 1
\\
 Tyc &  2267-721-1 &1
 \\
 SAO &53674 & 1
 \\
 Sp. Typ. & F8 & 1
 \\
 E(B-V) &0.057&2\\
 $A_v$&$0\fm180$&2\\
\hline
\end{tabular}
\\
$^1${SIMBAD},
$^2${NASA/IPAC:http://irsa.ipac.caltech.edu},
\end{center}
\end{table}

\begin{table}
\begin{center}
\caption{Data from Hipparcos and Tycho Catalogues.} \label{table2}
\begin{tabular}{l|c}\hline
  & Hip689
   \\
 & HD375
  \\
 \hline
  $V_J(Hip)$ & $7\fm41$
   \\
  $(B-V)_J(Hip)$ & $0\fm606\pm0.015$
  \\
  $B_T$ & $8\fm113\pm0.009$
  \\
 $V_T$ & $7\fm470\pm0.007$
 \\
 $(B-V)_J(Tyc)$ & $0\fm584\pm0.009$
 \\
 $\pi_{Hip}$ (mas) old & $12.72\pm0.86$
 \\
  $\pi_{Hip}^{*}$ (mas) new & $11.69\pm0.67$
  \\
 $\pi_{Tyc}$ (mas) & $4.10\pm5.20$
 \\
\hline
\end{tabular}\\
$*$\cite{2007aaa...474..653V}
\end{center}
\end{table}

 The magnitude difference between the two components $\triangle m=0\fm27\pm0.01$ is adopted as the average of all $\triangle m$ measurements under the speckle filters  $545nm/30$
 (see Table ~\ref{deltam1}) as the closest filters to the visual. This value was used as an input to the equations:
  {\begin{eqnarray}
\label{eq1}
\ m_v^a=m_v+2.5\log(1+10^{-0.4\triangle m})
\end{eqnarray}}
and
 { \begin{eqnarray}
\label{eq2}
\ m_v^b=m_v^a+{\triangle m}
\end{eqnarray}}
\noindent
Using the entire visual  magnitude of the system $m_v=7\fm41 $ (see Table~\ref{table2}), the preliminary individual visual magnitude $m_v$ for each
component is: $m_v^a=8\fm04$ and $m_v^b=8\fm31$.

\begin{table}
\begin{center}
\caption{Magnitude difference between the components of the
system along with filters used to obtain the observations. }
\label{deltam1}
\begin{tabular}{l|c|c}
\noalign{\smallskip}
\hline
\noalign{\smallskip}
   $\triangle m $& filter ($\lambda/\Delta\lambda$)& ref.  \\
\hline
\noalign{\smallskip}
 $0\fm04\pm0.39$ & $V_{Hp}: 550nm/40 $& \cite{1997yCat.1239....0E} \\
 $0\fm28\pm0.05$ & $545nm/30 $& \cite{2005AAA...431..587P}\\
 $0\fm31\pm0.05$ & $545nm/30$ & \cite{2002aaa...385...87B}\\
 $0\fm03\pm0.15$ & $648nm/41$ & \cite{2004AJ....127.1727H}\\
 $0\fm22\pm0.24$ & $2115/nm214$ &\cite{2002aaa...385...87B}\\
 $0\fm02\pm0.15$ & $503/nm40$  & \cite{2004AJ....127.1727H}\\
 $0\fm00\pm0.15$ & $701/nm12$  & \cite{2004AJ....127.1727H}\\
 $0\fm20\pm0.15$ & $648/nm41$  & \cite{2004AJ....127.1727H}\\
 $0\fm20\pm0.15$ & $600nm/30 $ &\cite{2006BSAO...59...20B} \\
 $0\fm81       $ & $698/nm39$  & \cite{2008AJ....136..312H}\\
 $0\fm23\pm0.06$ & $545nm/30 $ &\cite{2006BSAO...59...20B} \\
 $0\fm22\pm0.03$ & $600nm/30 $ &\cite{2006BSAO...59...20B} \\
 $0\fm01       $ & $745/nm44$  & \cite{2008AJ....136..312H}\\
 $0\fm41       $ & $550/nm40$  & \cite{2008AJ....136..312H}\\
 $0\fm15       $ & $541/nm88$  & \cite{2008AJ....136..312H}\\
 $0\fm55       $ & $698/nm39$  & \cite{2008AJ....136..312H}\\
 $0\fm47       $ & $650/nm38$  & \cite{2008AJ....136..312H}\\
 $0\fm41       $ & $650/nm38$  & \cite{2008AJ....136..312H}\\
 $0\fm49       $ & $698/nm39$  & \cite{2008AJ....136..312H}\\
 $0\fm54       $ & $745/nm44$  & \cite{2008AJ....136..312H}\\
 $0\fm29       $ & $550/nm40$  & \cite{2008AJ....136..312H}\\
 $0\fm20\pm0.04$ & $600nm/30 $ & \cite{2007AstBu..62..339B}\\
 $0\fm01       $ & $550/nm40$  & \cite{2008AJ....136..312H}\\
 $0\fm07       $ & $745/nm44$  & \cite{2008AJ....136..312H}\\
 $0\fm00       $ & $745/nm44$  & \cite{2008AJ....136..312H}\\
 $0\fm04       $ & $550/nm40$  & \cite{2011AJ....141..180H}\\
 $0\fm88       $ & $550/nm39$  & \cite{2011AJ....141..180H}\\
 $0\fm71       $ & $698/nm39$  & \cite{2011AJ....141..180H}\\
 $0\fm00       $ & $745/nm44$  & \cite{2010AJ....139..205H}\\
 $0\fm48       $ & $550/nm40$  & \cite{2010AJ....139..205H}\\
 $0\fm43$        & $692/nm40$  &\cite{2009AJ....137.5057H}\\
 $0\fm52$        & $562/nm40 $ &\cite{2009AJ....137.5057H}\\
 $0\fm38$        & $692/nm40 $ &\cite{2009AJ....137.5057H}\\
 $0\fm48$        & $447/nm60 $ &\cite{2009AJ....137.5057H}\\

 \hline
\noalign{\smallskip}
\end{tabular}
\\
\end{center}
\end{table}

Preliminary individual absolute magnitudes were calculated using equation \ref{eq3}, assuming that both components are main sequence stars. These were used to calculate the preliminary input elements ($T_{eff}^a=6750K , T_{eff}^b=6500K, \log g_{a}=4.19 \,\, \textrm{and} \, \log g_{b}=4.21$) to construct  model atmospheres for each component using  grids of  Kurucz's 1994
blanketed models (ATLAS9). Once needed, Equations \ref{eq4}~\& ~\ref{eq5} are used, interstellar reddening is taken from Table~\ref{table1}, $T_\odot=5777\rm{K}$ is used and bolometric corrections are taken from \cite{1992adps.book.....L} and \cite{2005oasp.book.....G}.
Hence a spectral energy distribution for each component are built.

\begin{eqnarray}
\label{eq3}
\ M_v=m_v+5-5\log(d)-A_v \\
\label{eq4}
\log(R/R_\odot)= 0.5 \log(L/L_\odot)-2\log(T/T_\odot) \\
\label{eq5}
\log g = \log(M/M_\odot)- 2\log(R/R_\odot) + 4.43
\end{eqnarray}

The total energy flux from a binary star is created from the net
luminosity of the components $a$ and $b$ located at a distance $d
$ from the Earth. One can write \citep{2007AN....328...63A}:

\begin{eqnarray}
\label{eq6}
   F_\lambda \cdot d^2 = H_\lambda ^a \cdot R_{a} ^2 + H_\lambda ^b
\cdot R_{b} ^2,
\end{eqnarray}
 \noindent from which

\begin{eqnarray}
\label{eq7}
 F_\lambda  = (R_{a} /d)^2(H_\lambda ^a + H_\lambda ^b \cdot(R_{b}/R_{a})^2) ,
\end{eqnarray}
\noindent
 where $H_\lambda ^a $ and  $H_\lambda ^b$ are the fluxes from a unit
surface of the corresponding component. $F_\lambda$ here
represents the entire SED of the system.

The resulting entire synthetic SED which is built using the preliminary input elements does not coincide with the observational one. It shows a lower color $(B-V)$ index, which means that the temperatures of the stars should be lower.

 Many attempts were made to achieve the best fit between the synthetic  SEDs and the observed  one. The preliminary calculated set is taken as starting values and  an iteration method for the different sets of elements is used. The best fit is evaluated using to the following criteria:
 \begin{itemize}
   \item The maximum values of the absolute flux (represent by the apparent magnitudes and calculated using synthetic photometry ).
   \item The inclination of the spectrum (represents by the color indices $(U-B)$, $(B-V)$ and $(v-b)$).
   \item The magnitude difference between the components ($ \triangle m$).
   \item The profiles of the absorption lines.
 \end{itemize}

 While the last three criteria depend mainly on $T_{eff}\, \textrm{and}\, \log g$, which were fulfilled using: $$T_{eff}^a=6100\pm50K,T_{eff}^b=5940\pm50K$$
$$ \log g_a=4.00\pm0.10, \log g_b=3.99\pm0.10,$$ the first criterion depends on the parallax of the system and the radii of the components (see equation \ref{eq7}). The estimated entire synthetic visual magnitudes according to the parallax of Hipparcos and the radii of  \cite{2005oasp.book.....G} (assuming that both components are main sequence stars) are higher (i.e. the absolute flux is lower) than the observed ones. This  means that either the parallax of the system is incorrect and the system is closer to earth or the system's components are no longer  main sequence stars but evolved and have higher radii.

\begin{figure}
\includegraphics[angle=0,width=8.5cm]{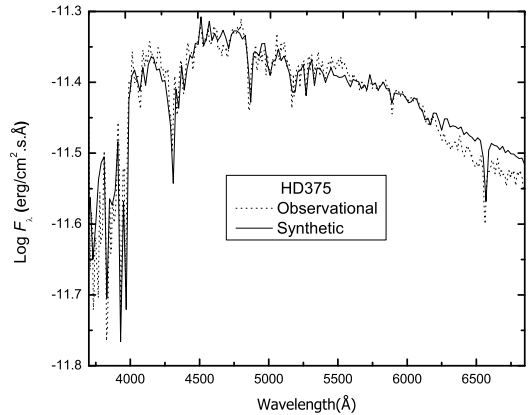}
\caption{Achieved best fit between the entire observational SED in the continuous spectrum of the system \citep{2002BSAO...53...58A} and the entire synthetic one of the two components built using Kurucz blanketed models \citep{1994KurCD..19.....K}.} \label{fit}
\end{figure}

Now, in order to get the exact fit with the observational absolute flux (Fig. \ref{fit}),  the parallax is chosen according to the following two approaches:
\begin{enumerate}
  \item Fixing the parallax as given by Hipparcos modified
      data $\pi=11.69\pm0.67$ mas
      \citep{2007aaa...474..653V}, and changing the radii
      till the best absolute flux reached. Note that while changing the radii, only slight changes in the value of $\triangle m$ are allowed.
  \item Fixing the radii as given by
      \cite{2005oasp.book.....G} tables or the standard
      R-L-T equation \ref{eq4} for the main sequence
      stars of $T_{eff}^a=(6100\pm50)K ,
      T_{eff}^b=(5940\pm50)K$ and changing the parallax
      till  the best absolute flux reached.

\end{enumerate}

Doubts in Hipparcos parallax measurements were introduced by \cite{1998AstL...24..673S}. They noted that,  in some cases, Hipparcos parallax measurements are distorted by the the orbital motion of the components of binary systems. Therefore, one has to be careful when using these measurements.

The first approach resulted in the following radii:
$$ R_a=2.00\pm0.15R_\odot, R_b=1.89\pm0.15R_\odot,$$ which refer to subgiant stars.

While the second approach resulted in the following radii and parallax:
$$ R_a=1.18\pm0.15 R_\odot, R_b=1.12\pm0.15R_\odot$$ and  $\pi=19.818$ mas ($
d=50.46\pm0.02 pc$), which disagrees with Hipparcos trigonometric parallax.

The estimated parallax obtained by the second approach does not coincide with orbital elements and mass sum calculated hereafter in this work (see Table \ref{orb}), while that given by Hipparcos was acceptable somehow.  Hence, the elements obtained   by the first approach represent the system better  than those obtained by the second approach, but not the best (see section \ref{Masses}).

\section {Orbital elements}
\label{Orbitalpar}
\begin{figure*}
\includegraphics[angle=0,width=16cm]{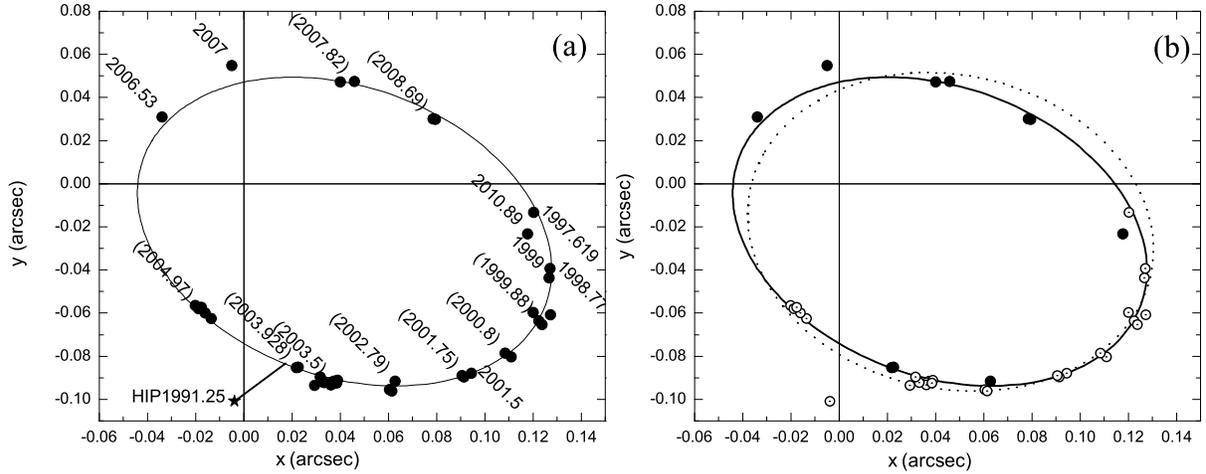}
 \caption{Relative visual orbit of the system HD375; The origin represents the position of the primary component. The filled circles are the new points used to modify the orbit (see Table ~\ref{obs.}) and  Hipparcos point is denoted by a star. (a) Shows the epoch of the positional measurements; Bracts mean that there is more than a point in that year. (b) Comparison between the modified orbit of this work (solid line) and that of  \cite{2008AJ....136..312H} (doted line). }
 \label{MH}
\end{figure*}

\begin{figure*}
\includegraphics[angle=0,width=16cm]{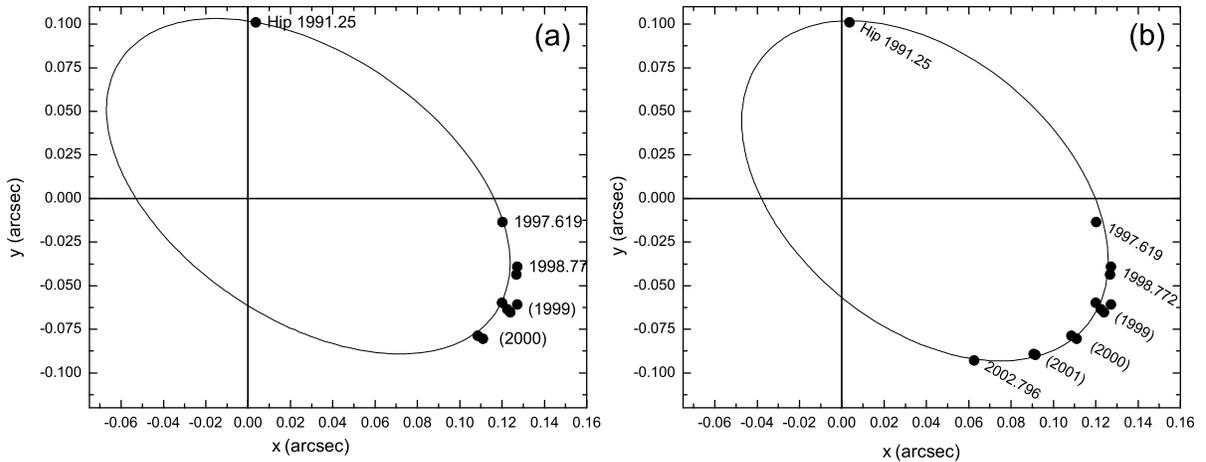}
 \caption{(a) The preliminary orbit of the system by \cite{2002aaa...385...87B}. (b) The orbit of \cite{2003PhDT.......174G}.}
 \label{MB}
\end{figure*}

The orbit of the system is built using the positional measurements listed in Table ~\ref{obs.}, which are taken from the Fourth Interferometric
Catalog and from \cite{2013AstBu..68...53B}. There are seven new points used to modify the orbit of \cite{2008AJ....136..312H}. Fig. \ref{MH}(a) shows the orbit of the system, which represents the relative positions of the secondary star with respect to the primary, and the ascending motion of the secondary according to the positional measurements. Fig. \ref{MH}(b) shows a comparison between the new orbit (solid line) and that of \cite{2008AJ....136..312H} (doted line). The preliminary orbit of \cite{2002aaa...385...87B}, and that of \cite{2003PhDT.......174G} are shown in Fig. \ref{MB}.
The modified orbital elements of the system along with the previous ones  are listed in Table ~\ref{orb}. It shows a good consistency between our estimated period, periastron epoch, semi-major axis and eccentricity and those estimated by \cite{2008AJ....136..312H}, while there are some differences in the inclination, position angle of nodes and the argument of periastron.

\section {Masses}
\label{Masses}
Using the estimated orbital elements, we calculated the total mass of the system (in  solar masses) and the corresponding error are calculated using the following equations:
\begin{eqnarray}
\label{eq9}
\ (M_a + M_b)/M_\odot=a^3/\pi^3 p^2 \\
\label{eq10}
\ \frac{\sigma_M}{M}=\sqrt{(3\frac{\sigma_\pi}{\pi})^2+(3\frac{\sigma_a}{a})^2+(2\frac{\sigma_p}{p})^2.}
\end{eqnarray}

The preliminary result using Hipparcos new trigonometric parallax $\pi$ (mas) =  $11.69\pm0.67$ is  $(M_a + M_b)/M_\odot=2.80\pm 0.49 $, while it is $2.19\pm 0.45 $ when using Hipparcos old trigonometric parallax $\pi$ (mas) =  $12.72\pm0.86$ (Table \ref{table2}).

  The calculated mass sum using Hipparcos new parallax gives higher value than what would be expected for two stars with the previously estimated physical elements, which is well enhanced by the positions of the two components on the evolutionary tracks. Another loop of iterated calculations is performed to reach the best fit between the estimated physical parameters and the orbital ones, especially the mass sum, which affected highly by the parallax value.

\begin{figure}
\includegraphics[angle=0,width=8.5cm]{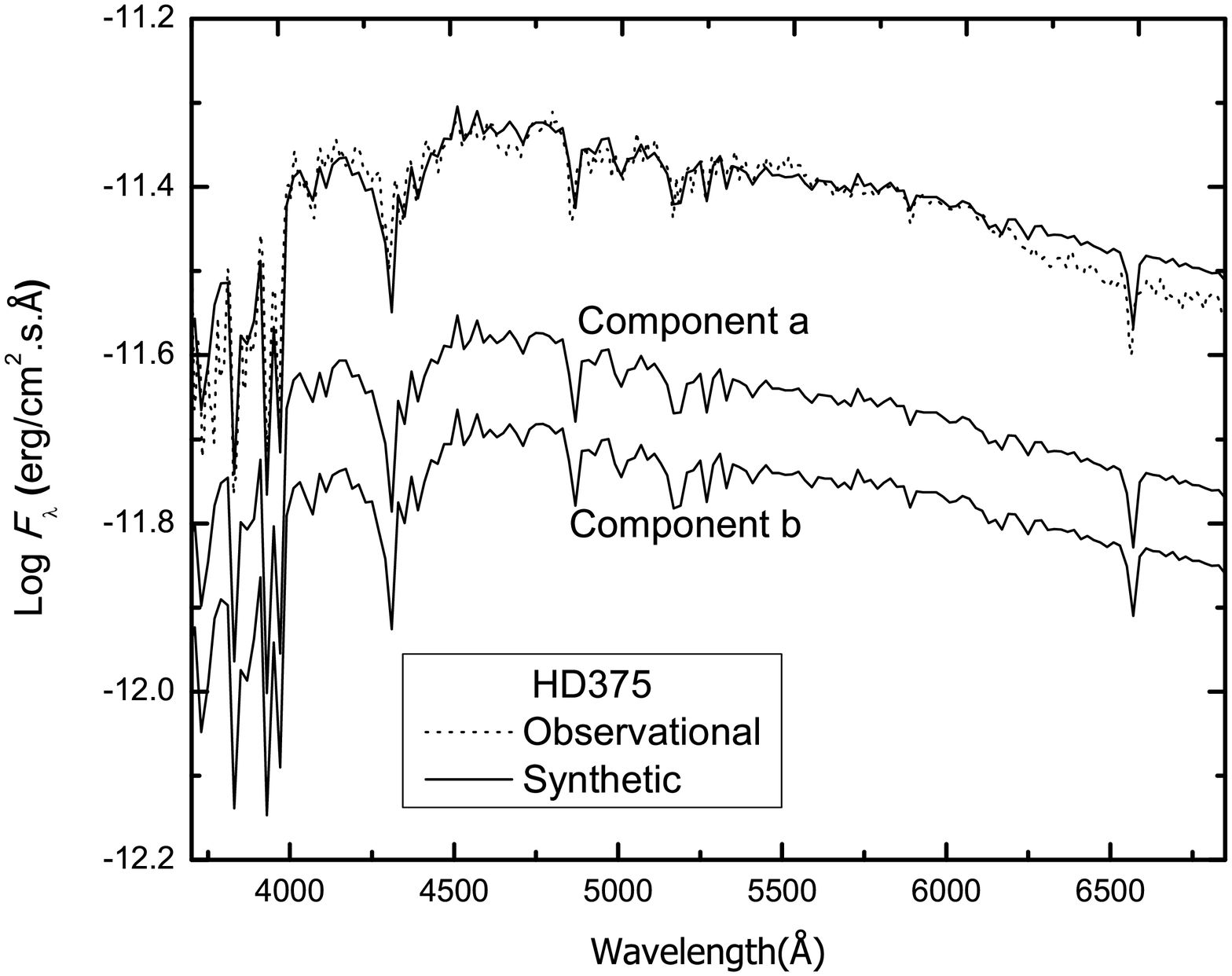}
\caption{Doted line: the entire observational SED in the continuous spectrum of the system. Solid lines: the  entire synthetic SED of the two components using a modified dynamical parallax ($\pi$ (mas) =  $12.02\pm0.60$); the synthetic flux of the primary component with $T_{\rm eff}=6100\pm50$\,K, log $g=4.01\pm0.10, R=1.93\pm0.15R_\odot$, and the synthetic flux of the secondary component with $T_{\rm eff} =5940\pm50$\,K, log $g=3.98\pm0.10, R=1.83\pm0.15 R_\odot $.} \label{Hip689all}
\end{figure}

The best fit (Fig.~\ref{Hip689all}) between the synthetic SED and the  observational one, along with the best consistency between  the physical and geometrical elements of both components,  dynamical parallax and dynamical mass sum are achieved using a modified dynamical parallax ($\pi$ (mas) =  $12.02\pm0.60$), which gives a mass sum of  $2.60\pm 0.16 $. The final physical and geometrical  elements of the system are listed in Table \ref{tablef1}, which adequately enough represent the  elements of the system within the error values of the measured quantities.

\begin{table}
\begin{center}
\caption{Positional measurements of the system from the Fourth Interferometric
Catalog and from \cite{2013AstBu..68...53B}.} \label{obs.}
\begin{tabular}{l|c|c|c}
\hline

 Epoch & $\theta\degr$ & $\rho\arcsec$ & Source \\
 \hline
  1991.25     & 358.0   &      0.101 	&  HIP1997
\\
  1997.6191   & 263.7 &        0.121 & Hor1999
\\
  1998.7717 &   72.9  &        0.133 &	Bag2002
\\
  1999.0145 &   71.0*   &      0.134 &	Hor2002
\\
  1999.7469 &   63.6  &        0.134  &	Bag2002
\\
  1999.8202 &   64.5*  &       0.141  & Msn2001
\\
  1999.8854 &   62.6 &         0.138 &	Hor2002
\\
  1999.8854 &   62.2 &         0.140 &	Hor2002
\\
  2000.7591 &   54.1  &        0.137 &	Hor2002
\\
  2000.8727 &   54.1 &         0.134 &	Bag2006
\\
  2001.4999  &   47.1*     &    0.129 &	Hor2008
\\
  2001.7526 &   45.6   &       0.128 &	Bag2006
\\
  2001.7526 &   45.5   &       0.127 &	Bag2006
\\
  2002.7879 &   32.4   &       0.113 &	Hor2008
\\
  2002.7879  &  32.6*   &      0.114 &	Hor2008
\\
  2002.796  &  34.4*   &      0.111 &	Bag2013
\\
  2003.5304 &   22.2* &        0.099 &	Hor2008
\\
  2003.5304&    21.5* &        0.099 &	Hor2008
\\
  2003.5305  &  23.1* &        0.099 &	Hor2008
\\
  2003.5305   & 21.2*   &      0.100 &	Hor2008
\\
  2003.6371   & 22.6*  &       0.100 &	Hor2008
\\
  2003.6371  &  19.8     &     0.098 &	Hor2008
\\
  2003.6371 &   17.5    &      0.098 &	Hor2008
\\
  2003.6371  &  19.5*   &      0.095 &	Hor2008
\\
  2003.928  &  14.8*   &      0.088 &	Bag2013
\\
  2003.928  &  14.4*   &      0.088 &	Bag2013
\\
  2004.8237   & 347.9   &      0.064 &	Bag2007
\\
  2004.9695  &  340.6   &      0.060 &	Hor2008
\\
  2004.9695  &  342.2*   &      0.061 &	Hor2008
\\
  2004.9723  &  345.2*   &      0.062 &	Hor2008
\\
  2004.9723  &  343.1 &        0.060 &	Hor2008
\\
  2006.5257  &  227.6* &  0.0457 &  Hor2011b
  \\
  2007.0106  &  185.4  & 0.0547  & Hor2011b
  \\
  2007.8172  &  139.8  &   0.062   &   Hor2010
  \\
  2007.8201  &  136.0  &  0.066  &  Hor2010
  \\
  2008.6910   & 111.0 &  0.084  & Hor2009
  \\
  2008.6937 & 110.7  &  0.085  & Hor2009
  \\
  2010.8919  &  78.9 &  0.12  &  Orl2011\\
\hline
\end{tabular}
\end{center}
${HIP1997a}${\cite{1997yCat.1239....0E}},
${Plz2005}${\cite{1999AJ....117..548H}},
${Bag2002}${\cite{2002AAA...385...87B}},
${Hor2004}${\cite{2004AJ....127.1727H}},
${Hor2002a}${\cite{2002AJ....123.3442H}},
${Msn2001b}${\cite{2001AJ....121.3224M}},
${Bag2006b}${\cite{2006BSAO...59...20B}},
${Hor2008}${\cite{2008AJ....136..312H}},
${Bag2013}${\cite{2013AstBu..68...53B}},
${Bag2007b}${\cite{2007AstBu..62..339B}},
${Hor2011b}${\cite{2011AJ....141..180H}},
${Hor2010}${\cite{2010AJ....139..205H}},
${Hor2009}${\cite{2009AJ....137.5057H}},
${Orl2011}${\cite{2011RMxAA..47..211O}},

* These points were modified by $180\degr$ to become consistent with the nearby points.
\end{table}

\begin{table*}
\begin{center}
\caption{Orbital elements of the system(\cite{2002aaa...385...87B}, \cite{2003PhDT.......174G}, \cite{2008AJ....136..312H} and this work)} \label{orb}
\begin{tabular}{l|c|c|c|c}
\hline

 Parameter & Balega et al. (2002) &  \cite{2003PhDT.......174G}& \cite{2008AJ....136..312H} & (this work)\\
  \hline

  $P$ (yr) & $19.3$ & $16.74\pm0.24$ &12.9 & $12.79\pm0.11$
   \\
  $T_o$ (yr)  & $2005.6$ & $1988.265\pm0.177$ & 2006.12 & $2006.36\pm0.02$
  \\
  $e$  & $0.38$ & $0.52\pm0.02$ & 0.6 & $0.5237\pm0.0051$
  \\
  $a$ (arcsec) & $0.124$ & $0.127\pm0.003$ & 0.091 & $0.0904\pm0.0005$
 \\
 $i$  (deg) & $125 $ & $ 124\pm2.0 $ & 159 & $149.03\pm1.13$
 \\
 $\Omega$ (deg) & $42$ & $ 32\pm3.0$ & 315 & $62.99\pm3.03$
\\
 $\omega$ (deg) & $107$ & $105\pm1.0 $ & 72 & $183.42\pm3.22$
\\
$(M_a + M_b)/M_\odot$ & $2.3^*$ & $3.55^*$ & $ 2.835^{**}$ & $2.83\pm 0.49^{**} $\\
                      &     &        &           & $2.19\pm 0.45 ^\dag$\\
                      &     &        &           & $2.60\pm 0.16 ^\ddag$\\
\hline
\end{tabular}
\end{center}
\flushleft{$^*$ \, Depending on the estimated individual absolute magnitudes supposing that both components are main sequence stars.\\
$^{**}$ Using Hipparcos new trigonometric parallax $\pi$ (mas) =  $11.69\pm0.67$.\\
$^\dag$ \, Using Hipparcos old trigonometric parallax $\pi$ (mas) =  $12.72\pm0.86$.\\
$^\ddag$ \, Using the estimated parallax  in this work $\pi$ (mas) =  $12.02\pm0.60$.}

\end{table*}

\begin{table}
\small
\begin{center}
\caption{Physical and geometrical elements of the components of the system.}
\label{tablef1}
\begin{tabular}{l|c|c}
\noalign{\smallskip}
\hline
\noalign{\smallskip}
Component & a &  b  \\
\hline
\noalign{\smallskip}
$T_{\rm eff}$\,(K) & $6100\pm50$ & $5940\pm50$ \\
Radius (R$_{\odot}$) & $1.93\pm0.20$ & $1.83\pm0.20$ \\
$\log g$ & $4.01\pm0.10$ & $3.98\pm0.10$ \\
$L (L_\odot)$ & $4.63\pm0.80 $  & $3.74\pm0.70$\\
$M_{V}$ & $3\fm26\pm0.40$ & $3\fm51\pm0.50$\\
Mass, ($M_{\odot})$& $1.35\pm0.16$ & $1.25 \pm0.15$  \\
$\overline{\rho}(\overline{\rho}_\odot)$& $0.188\pm0.015$& $0.204\pm0.015$\\
Sp. Type$^*$ & F8.5 IV &G0 IV \\
\hline
\multicolumn{1}{l}{Parallax (mas) }    & \multicolumn{2}{|c}{$12.02 \pm 0.60$}\\
\multicolumn{1}{l}{$(M_a + M_b)/M_\odot$ }    & \multicolumn{2}{|c}{$2.60\pm 0.16 $}\\
\multicolumn{1}{l}{Age$^*$ (Gy)}    & \multicolumn{2}{|c}{$3.5\pm 0.5 $}\\
\hline
\noalign{\smallskip}
\end{tabular}
\end{center}
$^*$Depending on the positions of the components on the evolutionary tracks of \cite{2005oasp.book.....G}.
\end{table}

\section{Synthetic photometry}
The following relation is used in the calculations of the entire and individual synthetic magnitudes of the system \citep{{2006AJ....131.1184M},{2007ASPC..364..227M}}:
\begin{equation}
m_p[F_{\lambda,s}(\lambda)] = -2.5 \log \frac{\int P_{p}(\lambda)F_{\lambda,s}(\lambda)\lambda{\rm d}\lambda}{\int P_{p}(\lambda)F_{\lambda,r}(\lambda)\lambda{\rm d}\lambda}+ {\rm ZP}_p\,,
\end{equation}
 where $m_p$ is the synthetic magnitude of the passband $p$, $P_p(\lambda)$ is the dimensionless sensitivity function of the passband $p$, $F_{\lambda,s}(\lambda)$ is the synthetic SED of the object and $F_{\lambda,r}(\lambda)$ is the SED of the reference star (Vega).  Zero points (ZP$_p$) from  \cite{2007ASPC..364..227M} (and references there in) are adopted.

The results of the calculated magnitudes and color  indices (Johnson-Cousins: $U, B, V, R, U-B, B-V, V-R $; Str\"{o}mgren: $u, v, b,
y, u-v, v-b, b-y$ and Tycho: $B_T, V_T, B_T-V_T$) of the entire system and individual components, in different photometrical systems,  are shown in Tables~\ref{synth1}.

\begin{table}
\small
\begin{center}
\caption{ Magnitudes and color indices  of the synthetic
spectra of the  system.} \label{synth1}
\begin{tabular}{l|c|c|c|c}
\noalign{\smallskip}
\hline
\noalign{\smallskip}
System & Filter & Entire & Comp. a& Comp. b\\
          &     &$\pm0.02$ &         \\
\hline
\noalign{\smallskip}
Johnson  & $U$    & 8.14 & 8.73 & 9.08 \\
 Cousins          & $B$ & 8.02   &  8.63 &  8.93  \\
               & $V$ & 7.41 &  8.04 &  8.29 \\
               & $R$ & 7.07 &  7.72 & 7.95  \\
               &$U-B$& 0.12 & 0.10 & 0.15 \\
               &$B-V$&0.61  &  0.59 &  0.64 \\
               &$V-R$& 0.33 &  0.32 & 0.35 \\
  \hline
\noalign{\smallskip}
  Str\"{o}mgren        & $u$ & 9.29 & 9.88 &  10.23  \\
                    & $v$ & 8.35 & 8.96  & 9.28  \\
                    & $b$ & 7.75 & 8.37 &  8.65 \\
                    &  $y$& 7.38 & 8.01 &  8.26  \\
                    &$u-v$& 0.94 &0.93& 0.95 \\
                    &$v-b$& 0.60 & 0.58 & 0.63 \\
                    &$b-y$& 0.37 & 0.36& 0.39 \\
  \hline
\noalign{\smallskip}
  Tycho       &$B_T$  & 8.17   & 8.77 & 8.09   \\
              &$V_T$  & 7.47   &8.11 & 8.36  \\
              &$B_T-V_T$& 0.69 & 0.67& 0.73\\
\hline
\noalign{\smallskip}
\end{tabular}
\end{center}
\end{table}

A comparison between the  synthetic visible magnitudes and
their color indices  with the
observational ones  of the system (Tables~\ref{synth1}) shows
a  good consistency within the three photometrical systems Johnson-Cousins, Str\"{o}\-mgren and Tycho (see
Table~\ref{synth3})
\begin{table}
\small
\begin{center}
\caption{ Comparison between entire synthetic visible magnitudes and color indices of the  system with the entire
 ones calculated from the observational SED \citep{2008AstBu..63..361A}.} \label{synth3}
\begin{tabular}{l|c|c|c}
\noalign{\smallskip}
\hline
\noalign{\smallskip}
System & Fil. & entire synth. & entire obs.\\
&     &$\pm0.02$ &        $\pm0.02$ \\
\hline
\noalign{\smallskip}
Johnson-
 Cousins   & $B$ & 8.02 &  8.03 \\
        & $V$ & 7.41 &  7.43 \\
        &$B-V$&0.61  &  0.60 \\

  \hline
\noalign{\smallskip}
  Str\"{o}mgren
                    & $v$ & 8.35 & 8.35 \\
                    & $b$ & 7.75 & 7.80 \\
                    &$v-b$& 0.60 & 0.55\\
  \hline
\noalign{\smallskip}
  Tycho       &$B_T$  & 8.17   & 8.18 \\
              &$V_T$  & 7.47   &7.50 \\
              &$B_T-V_T$& 0.69 & 0.67\\
\hline \noalign{\smallskip}
\end{tabular}
\end{center}
\end{table}

Depending on the tables of \cite{2005oasp.book.....G} or using \cite{1992adps.book.....L} $Sp-T_{\rm
eff}$ empirical relation,  the spectral types  of the system's components can be estimated as F8.5 and G0 for the components {a} and {b} respectively.

\section {Results and discussion}
Atmospheric modeling and visual magnitude difference between the two components  along with the entire observational SED are used to build synthetic individual and entire SED's for the components of the VCBS HD375.
The  least-square fitting with weights inversely proportional to the squares of the positional measurements observational errors is used to modify the orbit of the system. Hence, the physical and geometrical elements of the VCBS HD375 are estimated, and the parallax of the system is modified.

Fig. \ref{Hip689all} shows the achieved best fit between the entire synthetic SED's and the  observational one. Where we can see a good consistency of the  maximum values of the absolute flux and the inclination of the spectrum.
There is also a good consistency between the  synthetic magnitudes and colors  and the observational ones  within the three photometrical systems Johnson-Cousins, Str\"{o}\-mgren and Tycho (Tables~\ref{synth3} \& ~\ref{synth3565}).  This consistency gives a good indication about the reliability of the  estimated elements of the individual components of the system, which are listed  in Table~\ref{tablef1}.

The estimated masses and radii can only be explained by assuming that the system is  a subgiant  binary system. Earlier calculations of the mass sum are listed in Table \ref{orb}, and \cite{2012aaa...546A..69M} calculated it in three different ways; using Kepler's law (called the dynamical mass $M_d$), using the mass-luminosity relation along with the observed photometry (photometric mass $M_{ph}$) and using the mass-spectrum relation along with the spectral classification (spectral mass $M_{sp}$). They found $M_d(M_\odot)=2.78\pm 0.89 $, $M_{ph}(M_\odot)=2.67$ and $M_{sp}(M_\odot)=1.10 $. The discrepancy between the dynamical and spectral mass estimations is possibly due to their assumption that both components are main-sequence stars, where they used Table VI of \cite{1981Ap&SS..80..353S}, and that the spectral mass represents the minimum mass of the system.

A deep look at the estimated physical and geometrical elements of the system (Table~\ref{tablef1}) shows that the secondary component is very similar to the  star $\beta$ Hydri (HIP 2021), which is a G2IV evolved subgiant with an age of about 6.5 - 7.0 Gyr  \citep{1998aaa...330.1077D, 2003aaa...399..243F}. \cite{2007ApJ...663.1315B} used high precision asteroseismology to measure the mean stellar density of $\beta$ Hydri as $\overline{\rho}(\overline{\rho}_\odot)=0.1803\pm0.0011$ and \cite{2007MNRAS.380L..80N} used interferometry to measure its angular diameter, where they estimated its physical elements as:  $T_{\rm eff}(K)=5872\pm44$, $R(\textrm{R}_{\odot}=1.814\pm0.017$), $\log g=3.952\pm0.005$ $L (\textrm{L}_\odot)=3.51\pm0.09 $ and  mass $M (\textrm{M}_{\odot})=1.07\pm0.03$.

The  primary component is also similar to the secondary component of the binary system Beta Leonis Minoris ($\beta$ LMi B) which is known as an F8 subgiant with mass $M (\textrm{M}_{\odot})=1.7\pm0.4$ and absolute magnitude $M_v=2\fm3$ \citep{2002aaa...391..647G}.

\begin{figure}
\includegraphics[scale=0.32, angle=0]{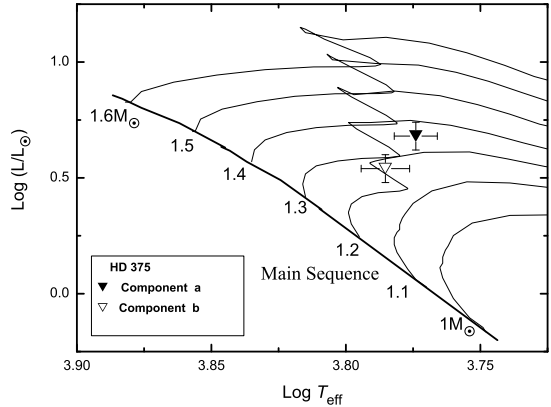}
\caption{The  systems' components  on the evolutionary tracks of  \cite{2000A&AS..141..371G}. }
\label{evol}
\end{figure}

That leads us to suggest that both components are evolved subgiant stars with age around 3.5 Gy. Fig.~\ref{evol} shows the positions of the components on the evolutionary tracks of  \cite{2000A&AS..141..371G}.

\begin{table}
\small
\begin{center}
\caption{ Comparison between the observational and synthetic
magnitudes, colors and magnitude differences of the system.} \label{synth3565}
\begin{tabular}{l|c|c}
\noalign{\smallskip}
\hline
\noalign{\smallskip}
HD375 & Obs.$^\dag$ & Synthetic (This work) \\
\hline
\noalign{\smallskip}
    $V_{J}$ & $7\fm41$ & $7\fm41$\\
   $B_T$  & $8\fm11\pm0.01$   &$8\fm17\pm0.02$\\
   $V_T$  & $7\fm47\pm0.01$   &$7\fm47\pm0.02$\\
   $(B-V)_{J}$&$ 0\fm61\pm0.02$ &$ 0\fm61\pm0.03$\\
   $\triangle m$  &$ 0\fm27^{\ddag}\pm0.01$  &$ 0\fm25\pm0.02$\\
\hline \noalign{\smallskip}
\end{tabular}\\
$\dag$ See Table~\ref{table2}\\
$\ddag$ Average value for the filter 545nm/30
(Table~\ref{deltam1}).
\end{center}
\end{table}

 Based on the similarity of both components, fragmentation
is proposed as the most likely formation process for the system.
Where \cite{1994MNRAS.269..837B} concludes that fragmentation of a rotating disk
around an incipient central protostar is possible, as long as
there is continuing infall.  \cite{2001IAUS..200.....Z} pointed out that
hierarchical  fragmentation during rotational collapse has been
invoked to produce binaries and multiple systems.

\section{Conclusions}

The  VCBS HDS375 was analyzed using atmospheric modeling and dynamical analysis.  The elements of the systems'  components  were estimated depending on the best fit between the entire observational SED  and  synthetic ones built using the atmospheric modeling of the individual components.
     The total and individual $UBVR$ Johnson-Cousins, $uvby$ Str\"{o}mgren and $BV$ Tycho synthetic magnitudes and colors of the system were calculated.

         A modified orbit and geometrical elements of the system were calculated and compared with earlier ones.
    Based on the estimated elements, especially radii and masses, we suggest that the two components are   F8.5~\&~G0 in their early subgiant stage, lying a bit upper the main-sequence on the H-R diagram. The estimated physical and geometrical elements of the two components coincide (within the error values) with those given by the Tables of \cite{1981Ap&SS..80..353S} for subgiants.

    Finally, fragmentation is proposed as the most
  likely process for the formation and evolution of both systems. Moreover, the system can  be used to test the stellar evolution theory and constraints on the physical description of the stellar interiors.

\section*{Acknowledgments}
 This work made use of SAO/NASA, SIMBAD, IPAC data systems and CHORIZOS code of photometric and spectrophotometric data  analysis. The authors express the  sincere thanks to Dr. Elliott P. Horch (Department of Physics, Southern Connecticut State University and  Kitt Peak National Observatory) for his critical comments.


\end{document}